\documentclass[pre,preprint,superscriptaddress,amsmath,amssymb,floatfix]
{revtex4}
\usepackage{graphics}
\usepackage[dvips]{graphicx}
\usepackage{amsfonts}
\usepackage{graphicx}
\usepackage[mathscr]{eucal}
\begin{document}
\title{Crossover behavior in fluids with Coulomb interactions}
 \author{O.V. Patsahan }
\address{Institute for Condensed Matter Physics of the National
Academy of Sciences of Ukraine, 1 Svientsitskii Str., 79011 Lviv, Ukraine}
 \author{J.-M. Caillol}
\address{Laboratoire de Physique Th\'eorique
CNRS UMR 8627, B\^at. 210
Universit\'e de Paris-Sud 91405 Orsay Cedex, France}
\author{I.M. Mryglod}
\address{Institute for Condensed Matter Physics of the National
Academy of Sciences of Ukraine, 1 Svientsitskii Str., 79011 Lviv, Ukraine}
 \date{\today}

\begin{abstract}
According to extensive experimental findings, the Ginzburg temperature $t_{G}$ for ionic fluids differs substantially from that of nonionic fluids [Schr\"oer W., Weig\"{a}rtner H. 2004 {\it  Pure Appl. Chem.} {\bf 76} 19]. A theoretical investigation of this outcome is proposed here by a mean field analysis of the interplay of  short and long range interactions on the value of  $t_{G}$. We consider a quite general continuous charge-asymmetric model made of charged hard spheres with  additional short-range interactions (without electrostatic interactions the model belongs to the same universality class as the $3D$ Ising model).
The effective Landau-Ginzburg Hamiltonian 
of the full system near its gas-liquid critical point is derived from which the Ginzburg temperature is calculated as a function of the ionicity. 
The results obtained in this way for $t_{G}$ are in good qualitative and sufficient quantitative agreement with available experimental data.
\end{abstract}

\maketitle
\section{Introduction}
It is known that electrostatic forces determine the properties of various systems: physical as well as chemical or biological. In particular, the Coulomb interactions are of great importance when dealing with ionic fluids i.e., fluids consisting  of dissociated cations and ions. In most cases the Coulomb interaction is the dominant interaction and due to its long-range character can substantially affect the critical properties and the phase behavior of ionic systems. Thus, the investigations concerning these issues are of great fundamental interest and practical importance.

Over the last ten years, both the phase  diagrams and the critical
behavior of  ionic solutions have been intensively studied using
both experimental and theoretical methods. These studies were
stimulated by  controversial experimental results, demonstrating
the three types of the critical behavior in electrolytes
solutions: (i) classical (or mean-field) and  (ii) Ising-like
behavior as well as  (iii) crossover between the two
\cite{singh_pitzer,levelt1,pitzer,gutkowski,Schroer-04,Schroer:review}. In
accordance with these peculiarities, ionic solutions were
conventionally divided into two classes, namely: ``solvophobic''
systems with Ising-like critical behavior in which  Coulomb forces are not supposed to play a major
role (the solvent is generally characterized by high dielectric
constant) and
``Coulombic'' systems in which the phase separation is primarily
driven by Coulomb interactions (the solvent is characterized by
low dielectric constant). Hence the criticality of the Coulombic systems became a challenge for theory and experiment. A theoretical model which demonstrates
the phase separation driven exclusively by Coulombic forces is a
restricted primitive model (RPM) \cite{fisher1,stell1}. In this
model the ionic fluid is described as an electroneutral binary
mixture of  charged hard spheres of equal diameter immersed in a
structureless dielectric continuum.  Early studies
\cite{stillinger,vorontsov,stellwularsen} established  that the model
has a gas-liquid phase transition. A reasonable theoretical
description of the critical point in the RPM was accomplished
at a mean-field (MF) level using integral equation methods
\cite{stell1,stell3} and Debye-H\"{u}ckel theory
\cite{levinfisher}. Due to controversial experimental findings,
the critical behavior of the RPM has  been under active debates
\cite{fisher3,schroer,Carvalho-Evans,caillol1,valleau,camp,luijten1,caillol_mc,patsahan_rpm, Patsahan-Mryglod-Caillol-05} and  strong evidence
for an Ising universal class has been found by recent  simulations
\cite{caillol_mc,luijten,kim:04:0}  and theoretical  \cite{ciach:00:0,
patsahan:04:1,ciach:05:0,ciach:06:1} studies.

In spite of significant progress in this field, the criticality of ionic systems are far from being completely understood. The investigation of more complex models is very important in understanding the nature of critical  behavior of real ionic fluids demonstrating both the charge and size asymmetry as well as other complexities such as  short-range attraction. A description of a  crossover region when the critical point is approached  is of particular interest for such models.
Based on the experimental findings one can suggest that in ionic fluids the temperature interval of crossover regime, characterized  by the Ginzburg temperature, is much smaller than observed in nonionic systems \cite{Schroer-04}. In particular, a sharp crossover was reported for the systems ${\rm Na-NH_{3}}$ \cite{Chieux-Sienko} (see also \cite{Narayanan-Pitzer1, Narayanan-Pitzer2, Anisimov}). The analysis of   experimental data for various ionic solutions confirmed that such systems generally exhibit crossover or, at least a tendency to crossover from the Ising behavior asymptotically close to the critical point, to the mean-field behavior upon increasing distance from the critical point \cite{Gutkowskii-Anisimov}.
Moreover, the systematic experimental investigations of the ionic systems such as  tetra-$n$-butylammonium picrate, ${\rm Bu_{4}NPic}$, (for tetra-$n$-butylammonium picrate we will follow the notations from  \cite{Schroer-04,Schroer:review}) in long chain $n$-alkanols with dielectric constant ranging from $3.6$ for $1$-tetradecanol to $16.8$ for $2$-propanol suggest an increasing  tendency for crossover to the mean-field behavior when the Coulomb contribution becomes essential \cite{Schroer-04,Schroer:review,Kleemeier}. They also indicate that the "Coulomb limit" reduced temperature of the RPM  $T_{c}\simeq 0.05$ is valid for the almost non-polar long chain alkanols \cite{Schroer:review,Kleemeier}. It has been stressed  \cite{Kleemeier} that  for solutions of ${\rm Bu_{4}NPic}$ in $1$-alkanols, the upper critical solution points are found to increase linearly with the chain length of the alcohols (that corresponds to the decrease of dielectric constant of the solvent). The experimental data for the critical points and the dielectric permittivities for solutions of  ${\rm Bu_{4}NPic}$ in $1$-alkanols are given in Table~1 \cite{Kleemeier}.
\begin{table}[htbp]
\caption{The experimental parameters of the critical points (critical temperature $T_{c}$,  critical mass fraction $w_{c}$) and the corresponding dielectric constants $\epsilon$ for solutions of  ${\rm Bu_{4}NPic}$ in $1$-alkanols \cite{Kleemeier}. }
\begin{center}
\begin{tabular}{|c|c|c|c|}
\hline Solvent & $\epsilon(T_{c})$& $T_{c}/K$ & $w_{c}$ \\
\hline $1$-oktanol & $9.5$& $298.55$& $0.336$  \\
\hline $1$-nonanol& $7.9$ &$308.64$&$0.325$ \\
\hline $1$-decanol& $6.4$ &$318.29$ & $0.3152$ \\
\hline $1$-undecanol & $5.4$ & $326.98$ & $0.303$  \\
\hline $1$-dodecanol & $4.7$ & $335.91$ & $0.2951$ \\
\hline $1$-tridecanol &$4.3$ & $342.35$& $0.284$ \\
\hline $1$-tetradecanol& $3.6$ &$351.09$&$0.2721$\\
\hline
\end{tabular}
\end{center}
\end{table}

Theoretically the crossover behavior in ionic systems was firstly studied   for the RPM \cite{fisher3,schroer,Carvalho-Evans}. The results obtained
for the Ginzburg temperature were similar to those found for
simple fluids in comparable fashion that is in variance to what is expected from the experiments \cite{Schroer-04,Schroer:review}. Nearly at the same time in
\cite{moreira-degama-fisher} the crossover behavior of the lattice
version of a fluid exhibiting the Ising behavior was studied as
additional symmetrical electrostatic interactions were turned on. Based on the
microscopic ground, the effective Hamiltonian in terms of the
fluctuating field conjugate to the number density was derived in
this work.  Then, the crossover between the mean-field and
Ising-like behavior was estimated using the Ginzburg criterion.
The resulting crossover temperature calculated as function of the
ionicity ${\cal I}$, which defines the strength of the Coulomb
interaction relative to the short-range interaction, indicates its
weak dependence but with the trends correlating with those observed experimentally.

In this paper we are also interested in the critical behavior of ionic fluids. 
In particular, we study the effect of the interplay of short-range  and long-range interactions on the crossover behavior in such systems. We consider  {\textit{a continuous version}} of the charge-asymmetric ionic fluid in
which both the long-range Coulomb  and  short-range van-der Waals-like interactions are included.
Following \cite{moreira-degama-fisher} we introduce  the ionicity 
\begin{equation}
{\cal I}=\frac{1}{\beta^{*}}=\frac{\vert q_{1}q_{2}\vert}{k_{B}T\epsilon\sigma},
\label{ionicity}
\end{equation}
where $q_{i}$ is the charge on ion $i$, $k_{B}$ is the Boltzmann constant, $T$ is the temperature, $\sigma$ is collision diameter and $\epsilon$ is the dielectric constant. Then we  derive the effective Hamiltonian of the charge-asymmetric model in the vicinity of the gas-liquid critical point. As in \cite{moreira-degama-fisher}, the  coefficients obtained for the effective Hamiltoninan have the forms of expansions in the ionicity but with new terms that appear in this case. Based on this Hamiltonian we estimate the Ginzburg temperatures as functions of the ionicity.

The layout of the paper is as follows. In Section~2 we introduce a continuous charge-asymmetric model with additional short-range attractive interactions included. We derive here the functional representation of the grand partition function of the model in terms of the fluctuating fields $\varphi_{\mathbf k}^{S}$ and $\varphi_{\mathbf k}^{D}$ conjugate to the total density and charge density, respectively. Section~3 is devoted to the derivation of the effective GLW Hamiltonian in the vicinity of the critical point. In Section~4 we calculate the Ginzburg temperature as a function of the ionicity for  different values of the range of the attractive potential. We conclude in Section~5. 

\section{Background}

\subsection{Model}
Let us start with a general case of a classical charge-asymmetric two-component system consisting of $N$ particles among which there are $N_{1}$ particles of species $1$ and $N_{2}$ particles of species $2$. The pair interaction potential is assumed to be of the following form:
\begin{equation}
U_{\alpha\beta}(r)=\phi_{\alpha\beta}^{HS}(r)+
\phi_{\alpha\beta}^{SR}(r)+\phi_{\alpha\beta}^{C}(r),
\label{2.1a}
\end{equation}
where $\phi_{\alpha\beta}^{HS}(r)$ is the interaction potential between the two  additive hard spheres of diameters $\sigma_{\alpha}$ and $\sigma_{\beta}$. We call the two-component hard sphere system a reference system. Thermodynamic
and structural properties of the reference system are assumed to be known. $\phi_{\alpha\beta}^{SR}(r)$ is the potential of the short-range (van-der-Waals-like ) attraction.
$\phi_{\alpha\beta}^{C}(r)$ is the Coulomb potential: $\phi_{\alpha\beta}^{C}(r)=q_{\alpha}q_{\beta}\phi^{C}(r)/\epsilon$, where $\phi^{C}(r)=1/r$ and $\epsilon$ is the dielectric constant. The solution is made of both positive and negative ions so that the electroneutrality condition is satisfied,i.e.
$\sum_{\alpha=1}^{2}q_{\alpha}c_{\alpha}=0$,
where $c_{\alpha}$ is the concentration of the species $\alpha$, $c_{\alpha}=N_{\alpha}/N$.
The ions of the species $\alpha=1$ are characterized by their hard sphere diameter $\sigma_{1}$ and their electrostatic charge $+q_{0}$ and those of species $\alpha=2$, characterized by diameter $\sigma_{2}$, bear opposite charge $-zq_{0}$ ($q_{0}$ is elementary charge and $z$ is the parameter of charge asymmetry). In general, the two-component system of hard spheres interacting via the potential $\phi_{\alpha\beta}^{SR}(r)$ can exhibit both the gas-liquid and demixion critical points which belong to the $3D$ Ising model universal class.

We consider the grand partition function (GPF) of the system which can be written as follows:
\begin{equation}
\Xi[\nu_{\alpha}]=\sum_{N_{1}\geq 0}\sum_{N_{2}\geq
0}\prod_{\alpha=1,2}
\frac{\exp(\nu_{\alpha}N_{\alpha})}{N_{\alpha}!} \int({\rm
d}\Gamma) \exp\left[-\frac{\beta}{2}\sum_{\alpha\beta}\sum_{ij}
U_{\alpha\beta}(r_{ij})\right].
\label{2.1}
\end{equation}
Here the following notations are used:
$\nu_{\alpha}$ is the dimensionless chemical potential, $\nu_{\alpha}=\beta\mu_{\alpha}-3\ln\Lambda$, $\mu_{\alpha}$ is the chemical potential of the $\alpha$th species, $\beta$ is the reciprocal temperature,
$\Lambda^{-1}=(2\pi m_{\alpha}\beta^{-1}/h^{2})^{1/2}$ is the inverse de Broglie thermal wavelength; $(\rm d\Gamma)$ is the element of configurational space of the particles: $(\rm d\Gamma)=\prod_{\alpha}\rm d\Gamma_{\alpha}$, $\rm d\Gamma_{\alpha}={\rm d}{\mathbf r}_{1}^{\alpha}{\rm d}{\mathbf r}_{2}^{\alpha}\ldots{\rm d}{\mathbf r}_{N_{\alpha}}^{\alpha}$.

Let us introduce the operators $\hat\rho_{{\mathbf k}}^{S}$ and $\hat \rho_{{\mathbf k}}^{D}$
\[
\hat\rho_{{\mathbf k}}^{S}=\sum_{\alpha}\hat\rho_{{\mathbf k},\alpha}
\qquad \hat \rho_{{\mathbf
k}}^{D}=\sum_{\alpha}q_{\alpha}\hat\rho_{{\mathbf k},\alpha},
\]
which are  combinations of the Fourier transforms of the microscopic number density of the species  $\alpha$:  $\hat\rho_{{\mathbf k},\alpha}=\sum_{i}\exp(-{\rm i}{\mathbf
k}{\mathbf r}_{i}^{\alpha})$. In this case  the part of the Boltzmann factor entering eq. (\ref{2.1}) which does not include  hard sphere interactions can be presented as follows:
\begin{eqnarray}
&&\exp\left[-\frac{\beta}{2}\sum_{\alpha\beta}\sum_{i,j}(U_{\alpha\beta}(r_{ij})-\phi_{\alpha\beta}^{HS}
(r_{ij}))\right]=\exp\left[-\frac{\beta}{2V}\sum_{{\bf k}}(\tilde\phi_{SS}(k)\hat\rho_{{\mathbf k}}^{S}\hat\rho_{{\mathbf -k}}^{S}\right.  \nonumber \\
&&
\left.+\tilde\phi_{DD}(k)\hat \rho_{{\mathbf k}}^{D}\hat \rho_{{\mathbf -k}}^{D}+ 2\tilde\phi_{SD}(k)\hat\rho_{{\mathbf k}}^{S}\hat \rho_{{\mathbf -k}}^{D})+\frac{\beta}{2V}\sum_{\alpha}N_{\alpha}\sum_{{\mathbf
k}}(\tilde\phi_{\alpha\alpha}^{SR}(k)+q_{\alpha}^{2}\tilde\phi^{C}(k))\right],
\label{2.2}
\end{eqnarray}
where
\begin{eqnarray}
\tilde{\phi}_{SS}(k)&=&\frac{1}{(1+z)^{2}}\left[
z^{2}\tilde{\phi}_{11}^{SR}(k) + 2z \tilde{\phi}_{12}^{SR}(k)
+\tilde{\phi}_{22}^{SR}(k)\right]  \nonumber \\
\tilde{\phi}_{DD}(k)&=&\frac{1}{(1+z)^{2}}\left[
\tilde{\phi}_{11}^{SR}(k)
-2\tilde{\phi}_{12}^{SR}(k)+\tilde{\phi}_{22}^{SR}(k)\right]
+\tilde{\phi}^{C}(k) \nonumber \\
\tilde{\phi}_{SD}(k)&=&\frac{1}{(1+z)^{2}}\left[ z
\tilde{\phi}_{11}^{SR}(k)
+(1-z)\tilde{\phi}_{12}^{SR}(k)-\tilde{\phi}_{22}^{SR}(k)\right]
\label{2.3}
\end{eqnarray}
 with
$\tilde{\phi}_{\alpha\beta}^{X\ldots}(k)$ being a Fourier transform of the corresponding interaction potential defined by
\begin{equation*}
\tilde{\phi}_{\alpha\beta}^{X\ldots}(k)=\int_{V}\;{\rm d}{\mathbf r} {\phi}_{\alpha\beta}^{X\ldots}(r)\exp(-{\rm i}{\mathbf k}{\mathbf r}),
\quad
{\phi}_{\alpha\beta}^{X\ldots}(r)=\frac{1}{V}\sum_{\mathbf k} \tilde{\phi}_{\alpha\beta}^{X\ldots}(k)\exp({\rm i}{\mathbf k}{\mathbf r}).
\end{equation*}
Now we simplify our model assuming that
\begin{itemize}
\item The hard spheres will all be of the same diameter $\sigma_{\alpha}=\sigma$.
\item $\widetilde{\phi}_{++}^{SR}(k)=\widetilde{\phi}_{--}^{SR}(k)=\widetilde{\phi}_{+-}^{SR}(k)=\widetilde{\phi}^{SR}(k)$.
\end{itemize}
With these restrictions the uncharged system can only exhibit a gas-liquid critical point and a possible demixion is ruled out.

Taking into account the assumptions mentioned above we thus have
\[
\tilde{\phi}_{SS}(k)=\tilde{\phi}^{SR}(k)<0, \quad
\tilde{\phi}_{DD}(k)=\tilde{\phi}^{C}(k)>0, \quad
\tilde{\phi}_{SD}(k)\equiv 0.
\]
Finally it will be convenient to introduce the effective range $b_{SR}$ of short-range interactions through the relations
\begin{eqnarray}
\label{sr}
\widetilde{\phi}^{SR}(k) = \widetilde{\phi}^{SR}(0)
\left( 1 - \left(b_{SR} \; k\right)^{2} \right) + \mathcal{O}(k^{4}) \; .
\end{eqnarray}

\subsection{Functional representation of the grand partition function of an ionic model}

Let us take advantage of the properties of Gaussian functional integrals to rewrite
\begin{eqnarray*}
\exp\left(\frac{1}{2}\sum_{\mathbf{k}}\widetilde{w}_{S}(k)\hat\rho_{{\mathbf k}}^{S}\hat\rho_{{\mathbf -k}}^{S}\right)&=&\frac{1}{{\cal N}_{w_{S}}}\int ({\rm d}\varphi^{S})\exp\left( -\frac{1}{2}\sum_{\mathbf{k}}\left[ \widetilde{w}_{S}(k)\right] ^{-1}\varphi_{{\mathbf k}}^{S}\varphi_{{\mathbf -k}}^{S}\right. \nonumber \\
&&\left.+\sum_{\mathbf{k}}\hat\rho_{{\mathbf k}}^{S}\varphi_{{\mathbf k}}^{S}
\right),
\end{eqnarray*}
\begin{eqnarray*}
\exp\left(-\frac{1}{2}\sum_{\mathbf{k}}\widetilde{w}_{C}(k)\hat\rho_{{\mathbf k}}^{D}\hat\rho_{{\mathbf -k}}^{D}\right)&=&\frac{1}{{\cal N}_{w_{C}}}\int ({\rm d}\varphi^{D})\exp\left( -\frac{1}{2}\sum_{\mathbf{k}}\left[ \widetilde{w}_{C}(k)\right] ^{-1}\varphi_{{\mathbf k}}^{D}\varphi_{{\mathbf -k}}^{D}\right. \nonumber \\
&&\left.+{\rm i}\sum_{\mathbf{k}}\hat\rho_{{\mathbf k}}^{D}\varphi_{{\mathbf k}}^{D}
\right),
\end{eqnarray*}
with
\begin{displaymath}
{\cal N}_{w_{S}}=\int ({\rm d}\varphi^{S})\exp\left( -\frac{1}{2}\sum_{\mathbf{k}}\left[ \widetilde{w}_{S}(k)\right] ^{-1}\varphi_{{\mathbf k}}^{S}\varphi_{{\mathbf -k}}^{S}\right)
\end{displaymath}
\begin{displaymath}
{\cal N}_{w_{C}}=\int ({\rm d}\varphi^{D})\exp\left( -\frac{1}{2}\sum_{\mathbf{k}}\left[ \widetilde{w}_{C}(k)\right] ^{-1}\varphi_{{\mathbf k}}^{D}\varphi_{{\mathbf -k}}^{D}\right).
\end{displaymath}
and
\begin{displaymath}
({\rm d}\varphi^{A})=\prod_{{\mathbf k}}'{\rm d}\varphi_{{\mathbf k}}^{A}=
\prod_{{\mathbf k}}'{\rm d}(\Re \varphi_{{\mathbf k}}^{A}){\rm d}(\Im \varphi_{{\mathbf k}}^{A}), \qquad A=S,D.
\end{displaymath}

In the above equations we also introduced the notations $\widetilde{w}_{S}(k)=-\beta\widetilde{\phi}_{SS}(k)/V$ and $\widetilde{w}_{C}(k)=\widetilde{\phi}^{C}(k)/V$.

As a result, we can rewrite $\Xi[\nu_{\alpha}]$ in the form of a functional integral
\begin{equation}
\Xi[\nu_{\alpha}]=\frac{1}{{\cal N}_{w_{S}}}\frac{1}{{\cal N}_{w_{C}}}\int ({\rm d}\varphi^{S})({\rm d}\varphi^{D})\exp\left(-{\cal H}[\nu_{\alpha},\varphi^{S},\varphi^{D}] \right),
\label{2.5}
\end{equation}
where the action ${\cal H}$ reads as
\begin{eqnarray}
\label{action}
{\cal H}[\nu_{\alpha},\varphi^{S},\varphi^{D}]&=&
\frac{1}{2}\sum_{\mathbf{k}}\left[\widetilde{w}_{S}(k)\right] ^{-1}\varphi_{{\mathbf k}}^{S}\varphi_{-{\mathbf k}}^{S}+\frac{1}{2}\sum_{\mathbf{k}}\left[\widetilde{w}_{C}(k)\right] ^{-1}\varphi_{{\mathbf k}}^{D}\varphi_{-{\mathbf k}}^{D} \nonumber \\
&&-\ln\Xi_{HS}[\overline\nu_{S}+\varphi^{S},\overline\nu_{D}+{\rm i}\beta^{1/2}\varphi^{D}],
\end{eqnarray}

\begin{equation}
\overline\nu_{S}=\frac{z}{1+z}\bar\nu_{1}+\frac{1}{1+z}\bar\nu_{2}, \qquad
\bar\nu_{D}=\frac{1}{q_{0}(1+z)}(\bar\nu_{1}-\bar\nu_{2}).
\label{chem-pot}
\end{equation}
where the "renormalized" chemical potentials  $\overline{\nu}_{\alpha}$ are defined as
\begin{eqnarray}
\label{nu-bar}
\overline{\nu_{\alpha}}=\nu_{\alpha}+\frac{1}{2}\sum_{\mathbf{k}}\left(-\widetilde{w}_{S}(k)+\beta q_{\alpha}^{2}\widetilde{w}_{C}(k)\right) , \qquad \alpha=1,2.
\end{eqnarray}

Let us define
$\Delta\nu^{S}=\overline\nu_{S}-\varphi_{0}^{S}$
and
$\widetilde\varphi_{\mathbf{k}}^{S}=\Delta\nu^{S}+\varphi_{\mathbf{k}}^{S}$
with $\varphi_{0}^{S}$ chosen as the chemical potential of the hard spheres.
This leads to the relation
\begin{equation}
\overline\nu_{S}+\varphi^{S}=\varphi_{0}^{S}+\widetilde\varphi^{S}.
\end{equation}
Now we present $\ln\Xi_{HS}[\ldots]$ in the form of a cumulant expansion
\begin{eqnarray}
\ln\Xi_{HS}[\ldots]&=&\sum_{n\geq 0}\frac{1}{n!}\sum_{i_{n}\geq 1}
\sum_{{\mathbf{k}}_{1},\ldots,{\mathbf{k}}_{n}}
{\mathfrak{M}}_{n}^{(i_{n})}[\varphi_{0}^{S},\overline\nu_{D};k_{1},\ldots,k_{n}]
\widetilde\varphi_{{\bf{k}}_{1}}^{D}\ldots\widetilde\varphi_{{\bf{k}}_{i_{n}}}^{D}\nonumber \\
&&
\widetilde\varphi_{{\bf{k}}_{i_{n+1}}}^{S}\ldots\widetilde\varphi_{{\bf{k}}_{n}}^{S}\delta_{{\bf{k}}_{1}+\ldots
+{\bf{k}}_{n}},
\label{2.11}
\end{eqnarray}
where ${\mathfrak{M}}_{n}^{(i_{n})}[\varphi_{0}^{S},\overline\nu_{D};k_{1},\ldots,k_{n}]$ is the $n$th cumulant  (or the $n$th order truncated correlation function) defined by
\begin{equation}
{\mathfrak{M}}_{n}^{(i_{n})}[\varphi_{0}^{S},\overline\nu_{D};k_{1},\ldots,k_{n}]=\frac{\partial^{n}\ln
\Xi_{HS}[\ldots]}{\partial
\widetilde\varphi_{{\bf{k}}_{1}}^{D}\ldots\partial\widetilde\varphi_{{\bf{k}}_{i_{n}}}^{D}
\partial\widetilde\varphi_{{\bf{k}}_{i_{n+1}}}^{S}\ldots\partial\widetilde\varphi_{{\bf{k}}_{n}}^{S}}\vert_{\varphi_{0}^{S},\overline\nu_{D}}.
\label{2.12}
\end{equation}
In particular it follows from (\ref{2.12}) that
\begin{equation}
{\mathfrak{M}}_{0}^{(0)}=\ln\Xi_{HS}[\varphi_{0}^{S},\overline\nu_{D}].
\label{3.14}
\end{equation}
The expressions for the cumulants of higher order (for $i_{n}\leq 4$) are given in Appendix~A. It should be noted that, contrary to \cite{moreira-degama-fisher}, (\ref{2.11}) includes all powers  (even and odd) of the field $\varphi_{{\mathbf k}}^{S}$ conjugate to the total number density. It should be clear that the  coefficients in the cumulant expansion\ (12) depend on the  chemical potential (or, equivalently, on the density).

\section{Effective Hamiltonian in the vicinity of the critical point}

Taking into account (\ref{2.11}) we can rewrite (\ref{2.5})-(\ref{action}) as follows
\begin{eqnarray}
\Xi[\nu_{\alpha}]&=&\frac{1}{{\cal N}_{w_{S}}}\exp\left(-\overline{\cal H} \right)
\int ({\rm d}\widetilde\varphi^{S})\exp\left(-\frac{1}{2}\sum_{\mathbf{k}}\left[\widetilde{w}_{S}(k)\right] ^{-1}\widetilde\varphi_{{\mathbf k}}^{S}\widetilde\varphi_{-{\mathbf k}}^{S}
\right.\nonumber \\
&&
\left.+ \left[\widetilde{w}_{S}(0)\right] ^{-1}\Delta\nu^{S}\widetilde\varphi_{0}^{S}+\sum_{n\geq 1}\frac{1}{n!}
\sum_{{\mathbf{k}}_{1},\ldots,{\mathbf{k}}_{n}}
{\mathfrak{M}}_{n}^{(0)}[\varphi_{0}^{S},\overline\nu_{D}]\widetilde\varphi_{{\bf{k}}_{1}}^{S}\ldots\widetilde\varphi_{{\bf{k}}_{n}}^{S}\right.
\nonumber \\
&&
\left.\times\delta_{{\bf{k}}_{1}+\ldots
+{\bf{k}}_{n}}\right){\cal V}[\widetilde\varphi_{{\bf{k}}}^{S}],
\label{2.15}
\end{eqnarray}
where
\begin{equation*}
\overline{\cal H}=\frac{1}{2}\left[\widetilde{w}_{S}(0)\right] ^{-1}(\Delta\nu^{S})^{2}-\ln\Xi_{HS}[\varphi_{0}^{S}],
\end{equation*}

\begin{eqnarray}
{\cal V}[\widetilde\varphi_{{\bf{k}}}^{S}]&=&\frac{1}{{\cal N}_{w_{C}}}\int({\rm d}\widetilde\varphi^{D})\exp\left(-\frac{1}{2}\sum_{\mathbf{k}}\left[\widetilde{w}_{C}(k)\right] ^{-1}\widetilde\varphi_{{\mathbf k}}^{D}\widetilde\varphi_{-{\mathbf k}}^{D}+
\frac{1}{2}\sum_{\mathbf{k}}{\mathfrak{M}}_{2}^{(2)}\widetilde\varphi_{{\mathbf k}}^{D}\widetilde\varphi_{-{\mathbf k}}^{D}\right. \nonumber \\
&&
\left.+\frac{1}{2}\sum_{\mathbf{k}_{1},\mathbf{k}_{2},\mathbf{k}_{3}}{\mathfrak{M}}_{3}^{(2)}\widetilde\varphi_{{\mathbf k}_{1}}^{D}\widetilde\varphi_{{\mathbf k}_{2}}^{D}\widetilde\varphi_{{\mathbf k}_{3}}^{S}\delta_{{\bf{k}}_{1}+{\bf{k}}_{2}
+{\bf{k}}_{3}}+\frac{1}{4}\sum_{\mathbf{k}_{1},\ldots,\mathbf{k}_{4}}{\mathfrak{M}}_{4}^{(2)}\widetilde\varphi_{{\mathbf k}_{1}}^{D}\widetilde\varphi_{{\mathbf k}_{2}}^{D}\widetilde\varphi_{{\mathbf k}_{3}}^{S}\right.\nonumber \\
&&
\left.\times\widetilde\varphi_{{\mathbf k}_{4}}^{S}\delta_{{\bf{k}}_{1}+\ldots
+{\bf{k}}_{4}}+\frac{1}{6}\sum_{\mathbf{k}_{1},\ldots,\mathbf{k}_{4}}{\mathfrak{M}}_{4}^{(3)}\widetilde\varphi_{{\mathbf k}_{1}}^{D}\widetilde\varphi_{{\mathbf k}_{2}}^{D}\widetilde\varphi_{{\mathbf k}_{3}}^{D}\widetilde\varphi_{{\mathbf k}_{4}}^{S}\delta_{{\bf{k}}_{1}+\ldots
+{\bf{k}}_{4}}+\ldots
\right).
\label{2.16}
\end{eqnarray}
It is worth noting here that unlike to the case considered in
\cite{moreira-degama-fisher} we obtain in (\ref{2.16})  terms
proportional to
$\left(\widetilde\varphi^{S}\right)^{2}\left(\widetilde\varphi^{D}\right)^{2}$
and $\widetilde\varphi^{S}\left(\widetilde\varphi^{D}\right)^{3}$.
While the former is connected with an absence of a lattice symmetry, the 
the latter stems from  charge asymmetry.

Our aim now is to derive the effective Landau-Ginzburg (LG) Hamiltonian. Since we are interested in the gas-liquid critical point, this Hamiltonian should be written in terms of fields $\widetilde\varphi_{{\bf{k}}}^{S}$ conjugate to the fluctuation modes of the total number density.

To this end  we integrate out $\widetilde\varphi_{{\bf{k}}}^{D}$ in (\ref{2.16}) using a Gaussian measure. As a result, we can present ${\cal V}[\widetilde\varphi_{{\bf{k}}}^{S}]$ as follows:
\begin{eqnarray}
{\cal V}[\delta\widetilde\varphi_{{\bf{k}}}^{S}]&=&\frac{{\cal N}_{W_{C}}}{{\cal N}_{w_{C}}}\left[1+\langle {\cal A}\rangle_{G}+\frac{1}{2!}\langle {\cal A}^{2}\rangle_{G}+\frac{1}{3!}\langle {\cal A}^{3}\rangle_{G}+\ldots \right],
\label{2.17}
\end{eqnarray}
where $\langle\ldots\rangle_{G}$ means
\begin{displaymath}
\langle\ldots\rangle_{G}=\frac{1}{{\cal N}_{W_{C}}}\int ({\rm d}\varphi^{D})\;\ldots\exp\left(-\frac{1}{2}\sum_{\mathbf{k}}\widetilde{W}_{C}(k)\varphi_{{\mathbf k}}^{D}\varphi_{-{\mathbf k}}^{D}\right)
\end{displaymath}
with $\widetilde{W}_{C}(k)$ given by
\begin{equation}
\widetilde{W}_{C}(k)=\left[\widetilde{w}_{C}(k) \right]^{-1}+y^{2}{\widetilde G}_{1}
\label{2.18}
\end{equation}
and $y^{2}$ being the ionicity  introduced by (\ref{ionicity}): $y^{2}={\cal I}$

Taking into account (\ref{ionicity}) and the recurrence formulas of Appendix~A ${\cal A}$ may be written as a formal expansion in terms  of $y^{2}$
\begin{eqnarray}
{\cal A}&=&-\frac{y^{2}}{2}\sum_{\mathbf{k}_{1},\mathbf{k}_{2},\mathbf{k}_{3}}{\widetilde G}_{2}(k_{1},k_{2}+k_{3})\varphi_{{\mathbf k}_{1}}^{D}\varphi_{{\mathbf k}_{2}}^{D}\varphi_{{\mathbf k}_{3}}^{S}\delta_{{\bf{k}}_{1}+{\bf{k}}_{2}
+{\bf{k}}_{3}}-\frac{y^{2}}{4}\sum_{\mathbf{k}_{1},\ldots,\mathbf{k}_{4}}{\widetilde G}_{3}(k_{1},k_{2},k_{3}+k_{4})\nonumber \\
&&
\times\varphi_{{\mathbf k}_{1}}^{D}\varphi_{{\mathbf k}_{2}}^{D}\varphi_{{\mathbf k}_{3}}^{S}\varphi_{{\mathbf k}_{4}}^{S}\delta_{{\bf{k}}_{1}+\ldots
+{\bf{k}}_{4}}-\frac{{\rm i}y^{3}}{6}\frac{(1-z)}{\sqrt{z}}\sum_{\mathbf{k}_{1},\ldots,\mathbf{k}_{4}}{\widetilde G}_{2}(k_{1},k_{2}+k_{3}+k_{4})\nonumber \\
&&
\times\varphi_{{\mathbf k}_{1}}^{D}\varphi_{{\mathbf k}_{2}}^{D}\varphi_{{\mathbf k}_{3}}^{D}\varphi_{{\mathbf k}_{4}}^{S}\delta_{{\bf{k}}_{1}+\ldots
+{\bf{k}}_{4}}+\ldots.
\label{A}
\end{eqnarray}
In (\ref{2.18})-(\ref{A}) the ``tilde'' over $\varphi_{{\mathbf k}}^{D(S)}$ was omitted for the sake of simplicity.

It should be mentioned that the dependence of ${\widetilde G}_{n}(k_{1},k_{2},\ldots, k_{n})$ on the $k_{i}$ is very complicated. Since we consider  here the  behavior of the system near the critical point the limiting case of $k_{i}=0$ is of particular interest. Therefore, we substitute in (\ref{2.17})
\[
{\widetilde G}_{n}(k_{1},k_{2},\ldots, k_{n})\equiv{\widetilde G}_{n}(0,\ldots)
\qquad  n\geq 3
\]
and
\begin{equation}
{\widetilde G}_{2}(k)={\widetilde G}_{2}(0)(1+g^{2}k^{2}),
\label{G-2}
\end{equation}
with
\begin{equation}
g^{2}=\frac{\widetilde G_{22}(0)}{2\widetilde G_{2}(0)}, \qquad
\widetilde G_{22}(0)=\frac{\partial^{2}\widetilde G_{2}(k)}{\partial k^{2}}\vert_{k=0}.
\label{g-2}
\end{equation}

Having integrated out eq. (\ref{2.17}) $\Xi[\nu_{\alpha}]$ takes the form:
\begin{eqnarray*}
\Xi[\nu_{\alpha}]=\frac{1}{{\cal N}_{w_{S}}}\prod_{{\mathbf k}}\left(1+y^{2}\langle N\rangle_{HS}\widetilde w_{C}(k)\right)^{-1}\int\;({\rm d}\varphi^{S})\exp\left( -{\cal H}^{eff}[\varphi^{S}]\right),
\end{eqnarray*}

\begin{equation}
{\cal H}^{eff}[\varphi^{S}]=-\sum_{n\geq 0}\frac{1}{n!}\sum_{{\mathbf{k}}_{1},\ldots,{\mathbf{k}}_{n}}a_{n}\widetilde\varphi_{{\bf{k}}_{1}}^{S}\ldots\widetilde\varphi_{{\bf{k}}_{n}}^{S}\delta_{{\bf{k}}_{1}+\ldots+{\bf{k}}_{n}},
\end{equation}
where we have for the coefficients $a_{n}$ 
\begin{eqnarray}
a_{0}&=&-\overline{\cal H},\label{2.19a} \\
a_{1}&=&\langle N\rangle_{HS}+[\widetilde w_{S}(0)]^{-1}\Delta\nu^{S}-\frac{y^{2}}{2}\widetilde G_{2}(0)\sum_{{\mathbf q}}\widetilde\Delta(q)+\frac{y^{4}}{8}\left(3\widetilde G_{3}(0)+\frac{(1-z)^{2}-2z}{z}\right.\nonumber \\
&&
\left. \times\widetilde G_{2}(0) \right)\left[\sum_{{\mathbf q}}\widetilde\Delta(q)\right]^{2}, \label{2.19b} \\
a_{2}&=&-[\widetilde w_{S}(k)]^{-1}+\widetilde G_{2}(k)-\frac{y^{2}}{2}\widetilde G_{3}(0)\sum_{{\mathbf q}}\widetilde\Delta(q)+\frac{y^{4}}{2}[\widetilde G_{2}(0)]^{2}\sum_{{\mathbf q}}\widetilde\Delta(q)\widetilde\Delta(\mid{\mathbf k}+{\mathbf q}\mid), \label{2.19c} \\
a_{3}&=&\widetilde G_{3}(0)-\frac{y^{2}}{2}\widetilde G_{4}(0)\sum_{{\mathbf q}}\widetilde\Delta(q)+\frac{3}{2}y^{4}\widetilde G_{2}(0)\widetilde G_{3}(0)\sum_{{\mathbf q}}\widetilde\Delta^{2}(q), \label{2.19d} \\
a_{4}&=&\widetilde G_{4}(0)-\frac{y^{2}}{2}\widetilde G_{5}(0)\sum_{{\mathbf q}}\widetilde\Delta(q)+\frac{1}{2}y^{4}\left(3[\widetilde G_{3}(0)]^{2}+4\widetilde G_{2}(0)\widetilde G_{4}(0)\right)\sum_{{\mathbf q}}\widetilde\Delta^{2}(q), \label{2.19e}
\end{eqnarray}
and the propagator $\widetilde\Delta(q)$ is written as
\begin{eqnarray}
\widetilde\Delta(q)=\widetilde\Delta(q;y^{2})=[\widetilde W_{C}(q)]^{-1}=\frac{\widetilde w_{C}(q)}{1+y^{2}\langle N\rangle_{HS}\widetilde w_{C}(q)}.
\label{Delta}
\end{eqnarray}
Coefficients (\ref{2.19a})-(\ref{2.19e}) have the form of a formal expansion in terms of the ionicity ${\cal I}=y^{2}$. In our study all terms  which do not exceed the fourth order  of $y$ are kept. The ionicity is small enough for large values of the dielectric constant and increases with its decrease. From this point of view we can consider the expansions in (\ref{2.19a})-(\ref{2.19e}) for large values of $y^{2}$ only as formal ones. It should be also noted that $\widetilde\Delta(q) \sim 1/y^{2}$ (see (\ref{Delta})) for large values of $y^{2}$.

Let us introduce
\begin{equation}
r_{SR}=[\widetilde G_{2}(0)\widetilde w_{S}(0)]^{-1}-1=\frac{T-T_{c,0}}{T_{c,0}},
\label{r-sr}
\end{equation}
where $T_{c,0}=T_{c}({\cal I}=0)$ is the mean-field critical temperature of the uncharged system.

Taking into account  (\ref{r-sr}) we can rewrite $-{\cal H}^{eff}$ as follows:
\begin{eqnarray}
-{\cal H}^{eff}[\overline\varphi^{S}]&=&-\frac{1}{2}\left[\widetilde{w}_{S}(0)\right] ^{-1}(\Delta\nu^{S})^{2}+\ln\Xi_{HS}[\varphi_{0}^{S}]-\frac{1}{2}\sum_{\mathbf{k}}\left(r_{0} +\tau_{0}^{2}k^{2}\right) \overline\varphi_{{\mathbf k}}^{S}\overline\varphi_{-{\mathbf k}}^{S} \nonumber\\
&&
-\frac{v_{0}}{3!}\sum_{\mathbf{k}_{1},\mathbf{k}_{2},\mathbf{k}_{3}}\overline\varphi_{{\mathbf k}_{1}}^{S}\overline\varphi_{{\mathbf k}_{2}}^{S}\overline\varphi_{{\mathbf k}_{3}}^{S}\delta_{{\bf{k}}_{1}+{\bf{k}}_{2}
+{\bf{k}}_{3}}-\frac{u_{0}}{4!}\sum_{\mathbf{k}_{1},\mathbf{k}_{2},\mathbf{k}_{3},\mathbf{k}_{4}}\overline\varphi_{{\mathbf k}_{1}}^{S}\overline\varphi_{{\mathbf k}_{2}}^{S}\overline\varphi_{{\mathbf k}_{3}}^{S}\overline\varphi_{{\mathbf k}_{4}}^{S}\nonumber\\
&&
\times
\delta_{{\bf{k}}_{1}+{\bf{k}}_{2}
+{\bf{k}}_{3}+{\bf{k}}_{4}}-h_{0}\overline\varphi_{0}^{S},
\label{2.20}
\end{eqnarray}
where $\overline\varphi_{{\mathbf k}}^{S}=\sqrt{\widetilde G_{2}(0)}\varphi_{{\mathbf k}}^{S}$ and the following notations were introduced:
\begin{eqnarray}
r_{0}&=&r_{SR}+\frac{y^{2}}{2}\frac{\widetilde G_{3}(0)}{\widetilde G_{2}(0)}\sum_{{\mathbf q}}\widetilde\Delta(q)-\frac{y^{4}}{2}\widetilde G_{2}(0)\sum_{{\mathbf q}}\widetilde\Delta^{2}(q) \label{2.21a}\\
\tau_{0}^{2}&=&\tau_{SR}^{2}-\frac{y^{4}}{4}{\widetilde G_{2}(0)}\sum_{{\mathbf q}}\widetilde\Delta(q)\widetilde\Delta^{(2)}(q)\label{2.21b}\\
v_{0}&=&-\frac{1}{[{\widetilde G_{2}}]^{1/2}}\left(\frac{\widetilde G_{3}(0)}{\widetilde G_{2}(0)}-\frac{y^{2}}{2}\frac{\widetilde G_{4}(0)}{\widetilde G_{2}(0)}\sum_{{\mathbf q}}\widetilde\Delta(q)+
\frac{3y^{4}}{2}\widetilde G_{3}(0)\sum_{{\mathbf q}}\widetilde\Delta^{2}(q)\right) , \label{2.21c}\\
u_{0}&=&-\frac{1}{{\widetilde G_{2}}}\left(\frac{\widetilde G_{4}(0)}{\widetilde G_{2}(0)}-\frac{y^{2}}{2}\frac{\widetilde G_{5}(0)}{\widetilde G_{2}(0)}\sum_{{\mathbf q}}\widetilde\Delta(q)+
\frac{y^{4}}{2}\sum_{{\mathbf q}}\widetilde\Delta^{2}(q)\frac{1}{\widetilde G_{2}(0)}\right.
\nonumber \\
&&\left.\times\left[3[{\widetilde G_{3}(0)}]^{2}+4{\widetilde G_{2}(0)}{\widetilde G_{4}(0)}\right]\right), \label{2.21d} \\
h_{0}&=&-[\widetilde G_{2}(0)]^{1/2}\left(\frac{\langle N\rangle_{HS}+[\widetilde w_{S}(0)]^{-1}\Delta\nu^{S}}{\widetilde G_{2}(0)}-\frac{y^{2}}{2}\sum_{{\mathbf q}}\widetilde\Delta(q)+\frac{y^{4}}{8}\left[3\frac{\widetilde G_{3}(0)}{\widetilde G_{2}(0)}\right. \right.
\nonumber \\
&&\left.\left. +\frac{(1-z)^{2}-2z}{z}\right]\left[\sum_{{\mathbf q}}\widetilde\Delta(q)\right]^{2}\right), \label{2.21e}
\end{eqnarray}
where 
\begin{equation}
-\tau_{SR}^{2}=g^{2}+\frac{\bar b_{SR}^{2}}{\widetilde G_{2}(0)\left[\widetilde w_{S}(0)\right]^{2}}
\label{tau-sr}
\end{equation}
with
$\bar b_{SR}^{2}=b_{SR}^{2}\widetilde w_{S}(0)$ and $\widetilde\Delta^{(2)}(q)=\partial^{2}\widetilde\Delta(\mid{\mathbf k}+{\mathbf q}\mid)/\partial k^{2}\vert_{k=0}$.

Finally, we present (\ref{2.20}) as follows:
\begin{eqnarray}
-{\cal H}^{eff}[\overline\varphi^{S}]&=&-\frac{1}{2}\left[\widetilde{w}_{S}(0)\right] ^{-1}(\Delta\nu^{S})^{2}+\ln\Xi_{HS}[\varphi_{0}^{S}]-\frac{1}{2}\sum_{\mathbf{k}}\left(r +\tau^{2}k^{2}\right) \overline\varphi_{{\mathbf k}}^{S}\overline\varphi_{-{\mathbf k}}^{S} \nonumber\\
&&
-\frac{v}{\langle N\rangle_{HS}^{1/2}}\sum_{\mathbf{k}_{1},\mathbf{k}_{2},\mathbf{k}_{3}}\overline\varphi_{{\mathbf k}_{1}}^{S}\overline\varphi_{{\mathbf k}_{2}}^{S}\overline\varphi_{{\mathbf k}_{3}}^{S}\delta_{{\bf{k}}_{1}+{\bf{k}}_{2}
+{\bf{k}}_{3}}-\frac{u}{\langle N\rangle_{HS}}\sum_{\mathbf{k}_{1},\ldots,\mathbf{k}_{4}}\overline\varphi_{{\mathbf k}_{1}}^{S}\overline\varphi_{{\mathbf k}_{2}}^{S}\overline\varphi_{{\mathbf k}_{3}}^{S}\overline\varphi_{{\mathbf k}_{4}}^{S}\nonumber\\
&&
\times
\delta_{{\bf{k}}_{1}+{\bf{k}}_{2}
+{\bf{k}}_{3}+{\bf{k}}_{4}}-h\langle N\rangle_{HS}^{1/2}\overline\varphi_{0}^{S},
\label{2.22}
\end{eqnarray}
with
\begin{eqnarray*}
r=r_{0}, \qquad \tau^{2}=\tau_{0}^{2}, \qquad v=\frac{v_{0}}{3!}\langle N\rangle_{HS}^{1/2} \qquad
u=\frac{u_{0}}{4!}\langle N\rangle_{HS}, \qquad
h=h_{0}\langle N\rangle_{HS}^{-1/2}.
\end{eqnarray*}

At the critical point the following equalities hold
\[
r=0, \qquad
v=0, \qquad
h=0,
\]
which give the equations for the critical parameters i.e., the temperature, the density and the chemical potential at the critical point.

Equation (\ref{2.22}) gives the effective GLW Hamiltonian of the system (\ref{2.1a}) in the vicinity of the critical point. We are now in position to extract from eq.\ (\ref{2.22}) the Ginzburg temperature as a function of the ionicity.

Now let us specify  the short-range attraction, $\phi^{SR}(r)$, in the form of the square-well potential
\[
\phi^{SR}(r)=\left\{
\begin{array}{cc}  0, & ~~~~~~0 \leq r<\sigma\\
                  -\varepsilon, & ~~~~~~~ \sigma\leq r<\lambda\sigma\\

                                0, & ~~~ r\geq\lambda\sigma
\end{array}
\right. .~~~~~
\]
It is worth noting here that the system of hard spheres interacting through the potential $\phi^{SR}(r)$ with $\lambda=1.4-1.7$ reasonably models most simple fluids \cite{McQuarrie}. 
The Fourier transform of $\phi^{SR}(r)$  for the case of the Weeks-Chandler-Andersen (WCA) regularization inside the hard core \cite{wcha} has the form:
\begin{equation}
\widetilde \phi^{SR}(k)=\widetilde\phi^{SR}(0)\frac{3}{(\lambda x)^3}[-\lambda x
~\cos(\lambda x) + \sin(\lambda x)],
\label{phi-sr-k}
\end{equation}
where
$x=k\sigma$ and $
\widetilde \phi^{SR}(0)= -\varepsilon\sigma^3 \frac{4\pi}{3} \lambda^3$.

To be consistent we also use the WCA regularization scheme for the Coulomb potential which yields
\begin{equation}
\widetilde\phi^{C}(x)=4\pi\sin(x)/x^{3}.
\label{phi-c-k}
\end{equation}


\section{Ginzburg temperature}

Following \cite{moreira-degama-fisher} we can present the Ginzburg temperature by
\begin{equation}
t_{G}[\eta_{c}(y),y]\simeq \frac{18}{\pi^{2}}\frac{u^{2}(y)}{[1+t_{0}(y)]\tau^{6}(y)},
\label{temp-Ginzb}
\end{equation}
but in our case all quantities $u$, $t_{0}$ and $\tau^{2}$ should be estimated at critical density $\eta_{c}(y)$:
\[
u(y)=u(\eta_{c}(y),y),
\qquad
t_{0}(y)=t_{0}(\eta_{c}(y),y),
\qquad
\tau^{2}(y)=\tau^{2}(\eta_{c}(y),y).
\]
The density $\eta$ enters the expressions for $u$, $t_{0}$ and $\tau^{2}$ through
the structure factors $\widetilde S_{n}$.
A well-known criterium by Ginzburg predicts that the mean-field theory is valid only when $t_{G}<<\mid t\mid$ where
$t=\frac{T-T_{c}(y)}{T_{c}(y)}$ and $T_{c}(y)$ are the mean-field reduced temperature  and the mean-field critical temperature of the charged system at $\eta=\eta_{c}(y)$, respectively. 

In (\ref{temp-Ginzb}) $t_{0}(y)$ measures the increase of the mean-field temperature of the charged system in respect to the uncharged system
\begin{equation}
t_{0}(y)=\frac{T_{c}(y)}{T_{c,0}}-1
\label{t-0}
\end{equation}
which, for the model under consideration has the form:
\begin{equation}
t_{0}(y)=-\frac{y^{2}}{2}\frac{\widetilde G_{3}(0)}{\widetilde G_{2}(0)}\sum_{{\mathbf q}}\widetilde\Delta(q)+\frac{y^{4}}{2}\widetilde G_{2}(0)\sum_{{\mathbf q}}\widetilde\Delta^{2}(q).
\label{t-01}
\end{equation}

Taking into account that
\[
\left[\widetilde G_{2}(0)\widetilde w_{C}(0)\right]^{-1}=\frac{T}{T_{c,0}},
\]
and equation (\ref{t-0}), we can rewrite (\ref{tau-sr})  as follows:
\begin{equation}
-\tau_{SR}^{2}=g^{2}+\frac{b_{SR}^{2}}{\widetilde w_{C}(0)}(1+t_{0}(y)).
\label{tau-sr1}
\end{equation}

For the uncharged model the Ginzburg temperature reduces to
\[
t_{G}({\cal I}=0))=\frac{1}{32\pi^{2}}\frac{\widetilde S_{4}^{2}}{\widetilde S_{2}^{4}[\tau_{SR}({\cal I}=0)]^{6}},
\]
where $\widetilde S_{n}$ is given by (\ref{struct-factor}) and $\tau_{SR}({\cal I}=0)=\tau_{SR}(t_{0}=0)$.

First we calculate the critical density from the equation $v=0$. To this end we 
take into account (\ref{2.21c}), (\ref{phi-c-k}) and the formulas of Appendix~D. As a result, we obtain  the dependence of the dimensionless critical density
$\eta_{c}$ ($\eta=\pi\rho\sigma^{3}/6$) on the ionicity ${\cal
I}=y^{2}$ (see Fig.~1).
\begin{figure}
\centering
\includegraphics[height=7cm]{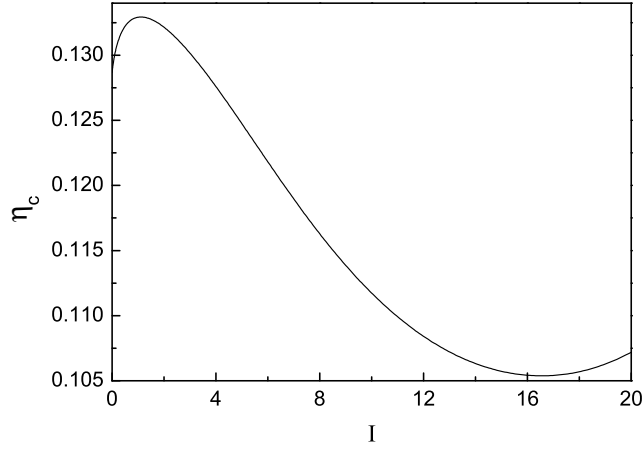}
\caption{Critical density as a function of ${\cal I}$.}
\end{figure}

In order to calculate the chemical potential at the critical point we introduce $\Delta\nu=\Delta\nu^{S}-\nu_{MF}^{S}$, where
\begin{equation*}
\nu_{MF}^{S}=-\langle N\rangle_{HS}\widetilde w_{S}(0)
\end{equation*}
is the mean-field value of the chemical potential $\overline \nu_{S}(0)$.
$\Delta\nu_{c}$ is obtained from the condition  $h=0$; taking into account (\ref{2.21e}) it yields  :
\begin{eqnarray}
\Delta\nu_{c}=\frac{y^{2}}{2}\widetilde S_{2}\sum_{{\mathbf q}}\widetilde\Delta(q)-\frac{y^{4}}{8}\left[3 \widetilde S_{3}+\frac{(1-z)^{2}-2z}{z}\widetilde S_{2}\right]\left[\sum_{{\mathbf q}}\widetilde\Delta(q)\right]^{2},
\label{dnu-c}
\end{eqnarray}
where
\begin{eqnarray}
\widetilde S_{n}(\eta_{c};0)=\widetilde G_{n}/\langle N\rangle_{HS}
\label{struct-factor}
\end{eqnarray}
is the $n$th particle structure factor at the critical density
$\eta_{c}({\cal I})$ when $k_{i}=0$. In Fig.~2 $\Delta\nu_{c}$ is
displayed as a function of the ionicity for different values of
the parameter $z$.
\begin{figure}
\centering
\includegraphics[height=8cm]{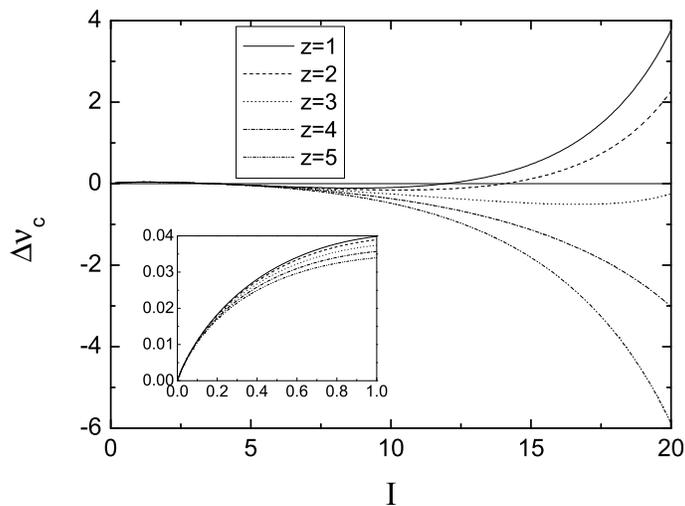}
\caption{$\Delta\nu_{c}$ as a function of ${\cal I}$ calculated from (\ref{dnu-c}) for different values of $z$ ($\eta=\eta_{c}$). The inset depicts the behavior of $\Delta\nu_{c}$ close to the origin.
}
\end{figure}

Now we calculate  $\tau^{2}$, $u$, $t_{0}$ and   $t_{G}$  at $\eta=\eta_{c}$ using (\ref{2.21b}), (\ref{2.21d}), (\ref{phi-sr-k})-(\ref{phi-c-k}), (\ref{t-01}) and formulas from Appendices~B-D.

The dependence of $\tau^{2}$ on ${\cal I}$ at different values of the parameter $\lambda$  is plotted  in Fig.~3. The explicit formula for $\tau_{SR}^{2}$ is given in Appendix~C. 
\begin{figure}
\centering
\includegraphics[height=7cm]{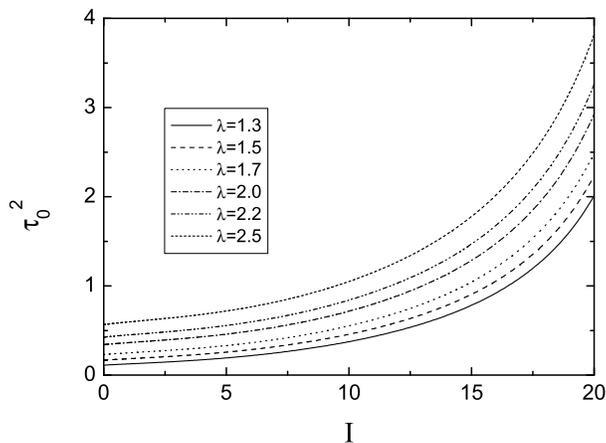}
\caption{The dependence of $\tau^{2}$ on the ionicity for different $\lambda$ ($\eta=\eta_{c}$).}
\end{figure}
The coefficient $u$ and the shift in the mean-field critical temperature, $t_{0}$, as functions of ${\cal I}$ are plotted in Figs.~4 and 5.  As is seen, quantities $\tau^{2}$, $t_{0}$ and $u$ are increasing functions of ${\cal I}$ in the whole region under consideration and their dependencies of ${\cal I}$ are at variance with those  obtained in \cite{moreira-degama-fisher} for the lattice model. Despite this fact, the behavior of the Ginzburg temperature as a function of ${\cal I}$ calculated in this work is qualitatively similar to that found in \cite{moreira-degama-fisher} (see Figs.~6-8). Moreover, as in \cite{moreira-degama-fisher}, the behavior of $t_{G}({\cal I})$ becomes nonmonotonic starting with some value of the  attraction potential range ($\lambda$ in our case). One can see in Fig.~7 that, for $\lambda=2$,  $t_{G}$    first drop off (at very small values of the ionicity) then increases slightly and at  ${\cal I}\simeq 1.23$ again starts to decrease.  In Fig.~8  the ratio of reduced Ginzburg temperatures, $t_{G}({\cal I})/t_{G}(0)$, is shown at different values of $\lambda$. It is worth noting that  the non-monotonic behavior of  $t_{G}({\cal I})$ becomes more pronounced as $\lambda$ increases.
\begin{figure}
\centering
\includegraphics[height=7cm]{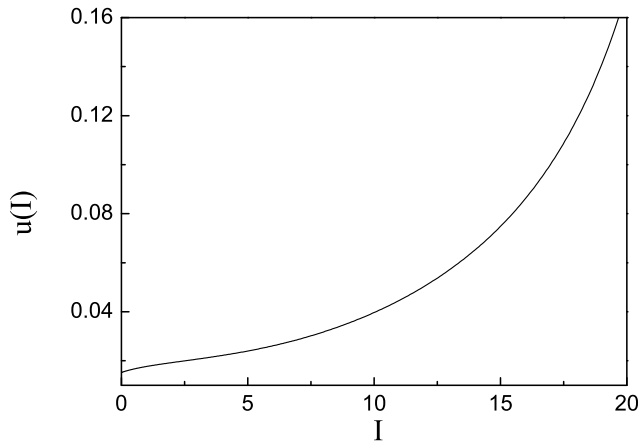}
\caption{The dependence of $u$ on the ionicity ${\cal I}$ ($\eta=\eta_{c}$).}
\end{figure}

\begin{figure}
\centering
\includegraphics[height=7cm]{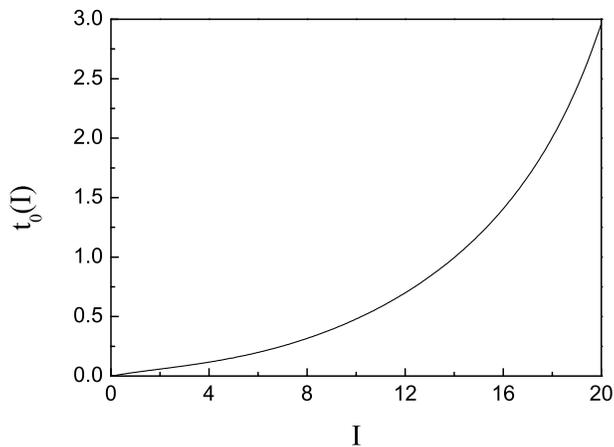}
\caption{The reduced shift of the mean-field critical temperature, $t_{0}$, as a function of ${\cal I}$ at $\eta=\eta_{c}$.}
\end{figure}
\begin{figure}
\centering
\includegraphics[height=7cm]{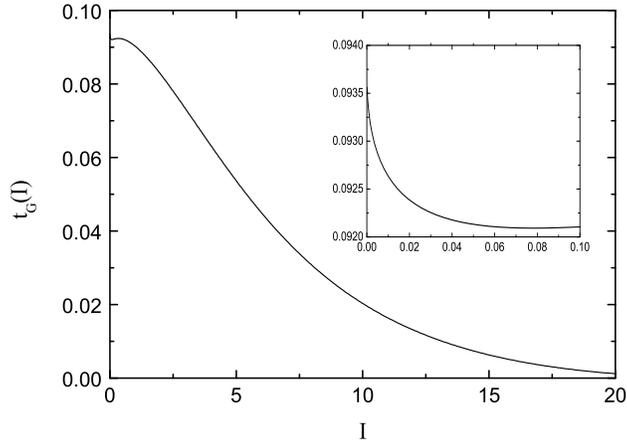}
\caption{The reduced Ginzburg temperature, $t_{G}$, as a function of ${\cal I}$ at $\lambda=1.5$ ($\eta=\eta_{c}$). The inset depicts the behavior of $t_{G}({\cal I})$ close to the origin.}
\end{figure}
\begin{figure}
\centering
\includegraphics[height=7cm]{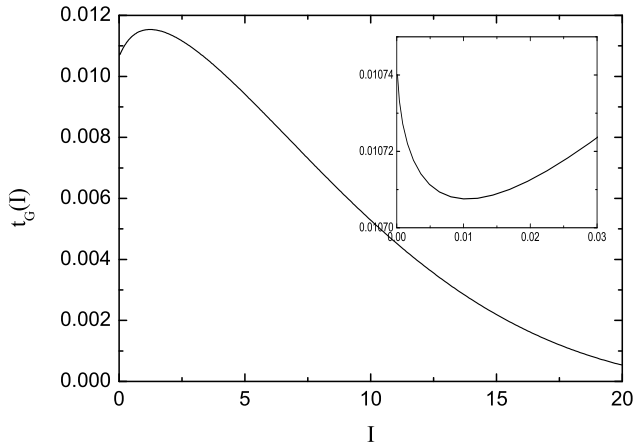}
\caption{The same as in Fig.~6 but at $\lambda=2$.}
\end{figure}
\begin{figure}
\centering
\includegraphics[height=7cm]{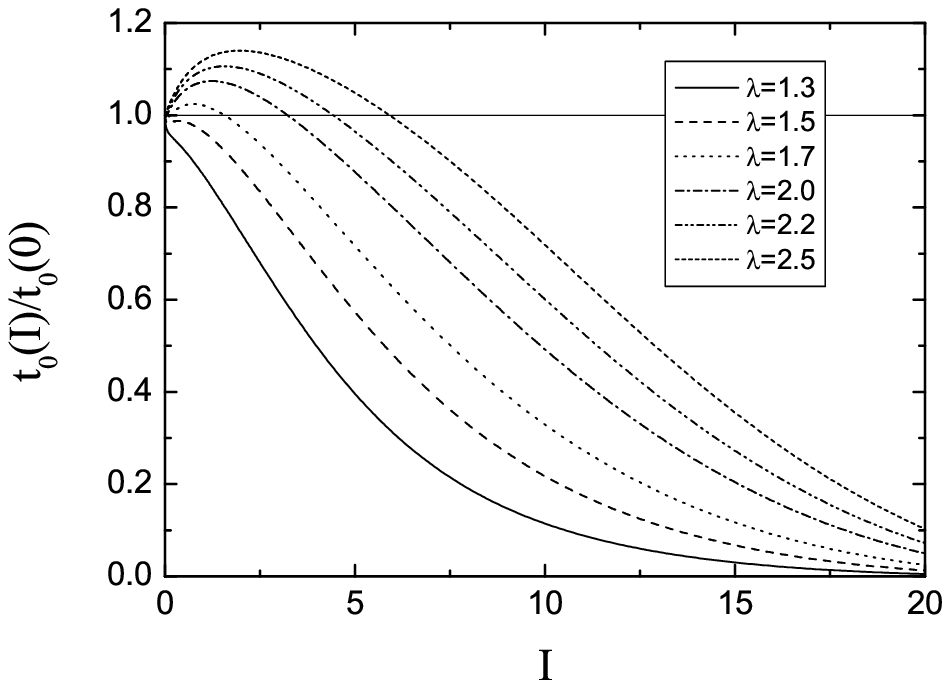}
\caption{The ratio $t_{G}(I)/t_{G}(0)$ as a function of the ionicity at different values of $\lambda$ ($\eta=\eta_{c}$).
}
\end{figure}

In Table~2 we compare our results for the ionicity dependence of the Ginzburg temperature (at  $\lambda=1.5$) with the results obtained in \cite {moreira-degama-fisher} for the lattice model as well as with experimental data for the crossover temperatures $t_{\times}$ (data for ${\cal I}$ and $t_{\times}$ are taken from \cite {moreira-degama-fisher}).
The systems (b)-(d) correspond to the same ionic species ${\rm Bu_{4}NPic}$ within  solvents of different dielectric constant. 
As is seen, in this case our results are in good agreement (qualitative and quantitative) with the experimental findings. The system (d) is ${\rm Na}$ in ${\rm NH_{3}}$ and, of course, might be described by the potential $\phi^{SR}(r)$ with the different attraction range $\lambda$. For instance, for $\lambda=2$ we obtain $t_{G}({\cal I}=6.97)=0.8\times 10^{-2}$ (see Fig.~7) that correlates with the experimental value $t_{\times}=0.6\times 10^{-2}$

\begin{table}[htbp]
\caption{Experimentally assessed crossover temperature, $t_{\times}$, taken from \cite {moreira-degama-fisher}: (a) tetra-$n$-butylammonium picrate (${\rm Bu_{4}NPic}$) in 1-tridecanol; (b) ${\rm Bu_{4}NPic}$ in 1-dodecanol; (c) ${\rm Bu_{4}NPic}$ in $75\%$ 1-dodecanol plus $25\%$ 1,4-butanediol; (d) ${\rm Na}$ in ${\rm NH_{3}}$;
(e) tetra-$n$-pentylammonium bromide in water and the  reduced Ginzburg temperature, $t_{G}$, found theoretically in \cite {moreira-degama-fisher} and in this work.}
\begin{center}
\begin{tabular}{|c|c|c|c|c|}
\hline System & Ionicity,${\cal I}$& $t_{\times}$ & $t_{G}$ (\cite {moreira-degama-fisher})& $t_{G}$ (this work)\\
\hline uncharged fluid & $0$& ${\cal O}({\cal I})$& $1$ & $\sim 0.09$ \\
\hline (a)& $17.9$ &$\sim 10^{-3}$&$\sim 0.712$&$2.7\times 10^{-3}$ \\
\hline (b)& $16.8$ &$\sim 0.9\times 10^{-2}$ & $\sim 0.717$& $0.38\times 10^{-2}$ \\
\hline (c) & $8.9$ & $\sim 3\times 10^{-2}$ & $\sim 0.777$ & $2.5\times 10^{-2}$ \\
\hline (d) & $6.97$ & $\sim 0.6\times 10^{-2}$ & $\sim 0.807$ & $3.7\times 10^{-2}$ \\
\hline (e) &$\sim 1.4$ & ${\cal O}({\cal I})$& $1$ & $\sim 0.09$\\
\hline
\end{tabular}
\end{center}
\end{table}

\section{Summary}
In this paper we study the reduced Ginzburg temperature as a function of the interplay between the short- and long-range interactions. The ionic fluid is modelled as  a charge asymmetric continuous system that includes additional short-range attractions. The model without  Coulomb interactions exhibits a gas-liquid critical point belonging to the Ising class of criticality. We derive an effective GLW Hamiltonian for the model whose coefficients  have the form of an expansion in  powers of the ionicity. Using these coefficients we calculate a Ginzburg temperature depending on the ionicity. To this end we introduce a specific model which consists of charged hard spheres of the same diameter interacting through the additional square-well  potentials. To study the effect of the interplay between short- and long-range interactions we change, besides the ionicity, the range of the square-well potential. 

As a result, we obtain the similar tendency for the reduced Ginzburg temperature as in \cite{moreira-degama-fisher} when the region of the short-range attraction increases i.e., its nonmonotonic character but with different numerical characteristics. However, our results  demonstrate a much faster decrease of the Ginzburg temperature when the ionicity increases. We found a good  qualitative  and sufficient quantitative agreement  with the experimental findings for ${\rm Bu_{4}NPic}$ in $n$-alkanols. This confirms the experimental observations that an interplay  between the solvophobic and Coulomb interactions alters the temperature region of the crossover regime i.e., the increase of the ionicity that can be related to the decrease of dielectric constant leads to the decrease of the crossover region. 
We suggest that the quantitative discrepancy of the results for $t_{G}$  obtained in \cite{moreira-degama-fisher} and in this work could be due to the fact, besides the difference in the symmetry of the two models,  that the chemical potential (or  density) dependence of the  Hamiltonian coefficients was taken into account explicitly in our case.

It should be noted that in the approximation considered in this paper only the critical chemical potential depends explicitly on the charge magnitude. In order to obtain the charge dependence of the other quantities terms of  order higher than $y^{2}$  should be taken into account
into the effective Hamiltonian.
Finally, we emphasize that  the functional representation (\ref{2.5})-(\ref{action}) allows to consider  more complicated models in particular models including charge and size asymmetry.

\section{Appendices}
\subsection{ Recurrence formulas for the cumulants  Fourier space.}
\begin{eqnarray*}
{\mathfrak{M}}_{n}^{(0)}(k_{1}, k_{2},\ldots, k_{n})&=& {\widetilde G}_{n}(k_{1},k_{2},\ldots, k_{n})
\nonumber \\
{\mathfrak{M}}_{n}^{(1)}(k_{1}, k_{2},\ldots, k_{n})&=&0 
\nonumber \\
{\mathfrak{M}}_{n}^{(2)}(k_{1}, k_{2},\ldots,k_{n})&=&
\left({\rm i}\right)^{2}\beta q_{\alpha}^{2}c_{\alpha}{\widetilde G}_{n-1}(k_{1}, k_{2},\ldots,\vert{\mathbf k}_{n-1}+{\mathbf k}_{n}\vert) 
\\
{\mathfrak{M}}_{n}^{(3)}(k_{1}, k_{2},\ldots, k_{n})&=&
\left({\rm i}\right)^{3}\beta^{3/2} q_{\alpha}^{3}c_{\alpha}{\widetilde G}_{n-2}(k_{1},k_{2},\ldots,\vert{\mathbf k}_{n-2}+{\mathbf k}_{n-1}+{\mathbf k}_{n}\vert) 
\\
{\mathfrak{M}}_{n}^{(4)}(k_{1}, k_{2},\ldots, k_{n})&=&\left({\rm i}\right)^{4}\beta^{2}\left\lbrace
3\left[q_{\alpha}^{2}c_{\alpha}\right] ^{2}{\widetilde G}_{n-2}(k_{1},k_{2},\ldots,\vert{\mathbf k}_{n-2}+{\mathbf k}_{n-1}+{\mathbf k}_{n}\vert) \right.  \nonumber \\
&&
\left.+\left(q_{\alpha}^{4}c_{\alpha}  -3\left[q_{\alpha}^{2}c_{\alpha}\right] ^{2}\right){\widetilde G}_{n-3}(k_{1}, k_{2},\ldots,\vert{\mathbf k}_{n-3}+\ldots+{\mathbf k}_{n}\vert)\right\rbrace ,
\end{eqnarray*}
where ${\widetilde G}_{n}(k_{1}, k_{2},\ldots, k _{n})$ is the Fourier transform of the $n$-particle  truncated correlation function \cite{stell} of a one-component hard sphere system and summation over repeated indices is meant.

\subsection{The nth-particle structure factors of a one component hard sphere systems in the Percus-Yevick approximation}
\begin{eqnarray}
S_{2}(0)&=&\frac{(1-\eta)^{4}}{(1+2\eta)^{2}},
\label{4b.1a} \\
S_{3}(0)&=&\frac{(1-\eta)^{7}(1-7\eta-6\eta^{2})}{(1+2\eta)^{5}},
\label{4b.1b} \\
S_{4}(0)&=&\frac{(1-\eta)^{10}(1-30\eta+81\eta^{2}+140\eta^{3}+60\eta^{4})}{(1+2\eta)^{8}},
\label{4b.1c} \\
S_{5}(0)&=&\frac{(1-\eta)^{13}(1-85\eta+957\eta^{2}-1063\eta^{3}-3590\eta^{4}-2940\eta^{5}-840\eta^{6})}{(1+2\eta)^{11}}
\label{4b.1d}
\end{eqnarray}

\subsection{Explicit expression for $\tau_{SR}^{2}$ }

Let us write the Ornstein-Zernike equation in the Fourier space
\begin{equation}
\tilde S_{2}(k)=\frac{1}{1-\rho \widetilde c(k)},
\label{S-2}
\end{equation}
where $\widetilde c(k)$ is the Fourier transform of the Ornstein-Zernike direct correlation function  \cite{hansen_mcdonald}
We have for $\widetilde c(k)$ in the Percus-Yevick
approximation \cite{ashcroft-1}
\begin{eqnarray}
\rho\widetilde c(k)&=&-24\eta\left(\alpha
k^{3}(\sin(k)-k\cos(k))+\beta
k^{2}(2k\sin(k)-(k^{2}-2)\cos(k)-2)\right.
\nonumber \\
&& \left. +\frac{1}{2}\eta\alpha((4k^{3}-24k)\sin(k)
-(k^{4}-12k^{2}+24)\cos(k)+24)\right)/k^{6}, \label{c-2}
\end{eqnarray}
where
\[
\alpha=\frac{(1+2\eta)^{2}}{(1-\eta)^{4}}, \qquad
\beta=-6\frac{\eta(1+\frac{1}{2}\eta)^{2}}{(1-\eta)^{4}}
\]
From (\ref{S-2}) and (\ref{c-2}) we get for $g^{2}$
\[
g^{2}=\frac{\widetilde G_{22}(0)}{2\widetilde G_{2}(0)}=0.05\eta\frac{(4\eta^{6}-27\eta^{5}+84\eta^{4}-146\eta^{3}+144\eta^{2}-75\eta+16)}
{(1+2\eta)^{2}(1-\eta)^{4}}.
\]
Taking into account (\ref{phi-sr-k}) we have $b_{SR}^{2}/\widetilde w_{S}(0)=0.1\lambda^{2}$.
As a result, $\tau_{SR}^{2}$  is as follows
\begin{equation}
\tau_{SR}^{2}=-0.05\left( \eta\frac{(4\eta^{6}-27\eta^{5}+84\eta^{4}-146\eta^{3}+144\eta^{2}
-75\eta+16)}
{(1+2\eta)^{2}(1-\eta)^{4}}+2\lambda^{2}(1+t_{0}(y))\right),
\label{tau-sr2}
\end{equation} 
where $t_{0}(y)$ is given by (\ref{t-01}).

\subsection{ Explicit expressions for the integrals used in equations (\ref{2.21a})-(\ref{2.21e})}

Using $\sum_{\mathbf k}=\frac{V}{(2\pi)^{3}}\int\, ({\rm d}{\mathbf k})$ we can present
\begin{eqnarray}
\sum_{{\mathbf k}}\widetilde\Delta(k)&=&\frac{2}{\pi}\int_{0}^{\infty}{\rm d}x\,x^{2}{\overline \Delta}(x), \label{i1} \\
\sum_{{\mathbf k}}\left( \widetilde\Delta(k)\right)^{2} &=&\frac{48\eta}{\pi\langle N\rangle_{HS}}\int_{0}^{\infty}{\rm d}x\,x^{2}\left( {\overline \Delta}(x)\right)^{2} , \label{i2} \\
\sum_{{\mathbf k}}\widetilde\Delta(k)\widetilde\Delta^{(2)}(k)&=&\frac{32\eta\sigma^{2}}{\pi\langle N\rangle_{HS}}\int_{0}^{\infty}{\rm d}x\,x{\overline \Delta}(x)\left( 2f_{1}(x)+xf_{2}(x)\right), \label{i3}
\end{eqnarray}
where the following notations are introduced:
\begin{eqnarray}
{\overline \Delta}(x)& =&\sin \left( x \right)\left( {x}^{3}+{\kappa^{*}}^{2}\,\sin \left( x \right)\right) ^{-1}, \label{delta-1}\\
f_{1}(x)&= &\left( {x}^{2} \left( \cos \left( x \right) x-3\,\sin \left( x \right)  \right) \right) \left( {x}^{6}+2\,{\kappa^{*}}^{2}{x}^{3}\,\sin \left( x \right)
+{{\kappa^{*}}}^{4}-{{\kappa^{*}}}^{4} \cos^{2} \left( x \right)\right) ^{-1}
 \label{f1}\\
f_{2}(x)&=&-x \left( {x}^{5}\sin \left( x \right) +{x}^{2}{\kappa^{*}}^{2}+{\kappa^{*}}^{2}{x}^{2}\, \cos^{2} \left( x \right)+6\,\cos \left( x \right) {x}^{4}-6\, {\kappa^{*}}^{2}x\,\sin \left( x \right)\cos \left( x \right) \right.
\nonumber \\
&&
\left.-12\,{x}^{3}\sin \left( x \right)+6\,{\kappa^{*}}^{2}-6\,{\kappa^{*}}^{2}\, \cos^{2} \left( x \right)\right) \left( {x}^{9}+3\,{\kappa^{*}}^{2}{x}^{6}\,\sin \left( x \right) +3\,{{\kappa^{*}}}^{4}{x}^{3} \right.
\nonumber \\
&&
\left.
-3\,{{\kappa^{*}}}^{4}{x}^{3} \cos^{2} \left( x \right)  +{{\kappa^{*}}}^{6}\sin \left( x \right) -{{\kappa^{*}}}^{6}\sin \left( x \right) \cos^{2}\left( x \right)\right) ^{-1}
\label{f2}
\end{eqnarray}
with $x=k\sigma$ and ${\kappa^{*}}=\kappa_{D}\sigma=\sqrt{24y^{2}\eta}$ being the reduced Debye number.

\end{document}